\documentclass[amssymb,showpacs,aps,prl,twocolumn]{revtex4}
\usepackage{epsfig}
\usepackage{amssymb,amsmath,amstext}
\begin{document}

\title{Perpendicular magnetic anisotropy in nanoclusters
 caused by surface segregation and shape}

\author{Stefan Heinrichs$^{1}$, Wolfgang Dieterich$^{1}$
 and Philipp Maass$^{2}$}

\affiliation{$^{1}$Fachbereich Physik, Universit\"at Konstanz,
             78457 Konstanz, Germany\\
$^{2}$Institut f\"ur Physik, Technische Universit\"at Ilmenau,
98684 Ilmenau, Germany}

\date{September 13, 2005}

\begin{abstract}
  The growth of binary alloy clusters on a weakly interacting
  substrate through codeposition of two atomic species is studied by
  kinetic Monte Carlo simulation. Our model describes salient features
  of CoPt$_3$-nanoclusters, as obtained recently by the molecular-beam
  epitaxy technique. The clusters display perpendicular magnetic
  anisotropy (PMA) in a temperature window of growth favorable for
  applications.  This temperature window is found to arise from the
  interplay of Pt surface segregation and aspect ratio for cluster
  shapes.  Conclusions are drawn how to optimize growth parameters
  with respect to PMA.
\end{abstract}

\pacs{81.15.Aa, 68.55.Ac, 75.75.+a}

\maketitle

Growth of frozen-in metastable structures by molecular beam
epitaxy (MBE) opens a wide field for producing materials
with novel properties. In most cases, the growth conditions require a
description based on far-from-equilibrium kinetics. Especially
interesting are multicomponent systems or systems with internal
degrees of freedom (magnetic or electrical moments) that couple to the
atomic structure. A rich variety of material properties is then
expected to emerge from the elementary processes of deposition, adatom
diffusion, surface nucleation, as well as from structural and
compositional changes of islands, from surface segregation or
magnetic ordering.

MBE growth of binary magnetic alloy thin films and nanoclusters has
recently attracted much attention in the search for materials with an
easy magnetic axis perpendicular to the substrate
\cite{Albrecht+05,Moser+02}. Such perpendicular magnetic anisotropy
(PMA) can be utilized to increase magnetic storage densities. In thin
magnetic films the dipolar form anisotropy favors alignment of the
magnetic moments parallel to the substrate. To achieve PMA in thin
films, the dipolar form anisotropy has to be overcome by effects based
on local atomic arrangements.

The system Co/Pt is particularly intriguing, as PMA could be obtained
in different ways: \emph{(i)} In Co/Pt multilayers the magnetic
moments of Co-atoms preferentially align along the interfacial Co-Pt
bonds and PMA is found in magnetic X-ray circular dichroism studies to
originate at the interfaces \cite{Nakajima+98}. \emph{(ii)} In
disordered films with CoPt$_3$-stoichiometry grown in [111] direction,
PMA is found in a temperature window 500~K $< T <$ 820~K
\cite{Shapiro+99}. XAFS \cite{Tyson+96} and EXAFS \cite{Cross+01}
experiments show the preferential occurrence of Co/Pt bonds out of
plane.  These were associated with surface-induced formation of flat
Co-rich domains, as suggested by the increase of Curie temperatures
with respect to those for disordered bulk Co/Pt-alloys
\cite{Shapiro+99}. \emph{(iii)} In CoPt$_3$ nanoclusters with 300 to
1000 atoms, grown by codeposition of Co and Pt on WSe$_2$
van-der-Waals surfaces, PMA is observed near ambient growth
temperatures \cite{Albrecht+01}, favorable for applications. The
underlying structural anisotropy, however, is not understood so far.

In fact, the occurrence of PMA in nanoclusters is surprising in view
of the kinetic limitations induced by comparatively low growth
temperatures. Several mechanisms can be envisaged in principle. One
possibility might be a bulk anisotropy arising from formation of
Co-rich domains in the bulk, similar to~\cite{Shapiro+99}.
Alternatively, PMA connected with the surface could be induced by
Pt surface segregation. A further question concerns the interplay
between L1$_{2}$ ordering and PMA.

In this Letter we investigate non-equilibrium structures that emerge
from the combined effects of deposition, surface diffusion and
ordering phenomena in a model for binary alloy growth.  We find that
surface segregation together with cluster shape formation leads to a
structural anisotropy that induces PMA. There is no evidence for
significant bulk contributions facilitating PMA.  In order to account
for the experimentally observed phenomena, exchange processes during
growth between low-coordinated atoms at the cluster surface have to be
included.

To demonstrate these mechanisms, we analyze a model based on nearest
neighbor interactions $V_{\rm AA}$, $V_{\rm AB}$, $V_{\rm BB}$ between
atomic species A and B, which are adjusted to the CoPt$_{3}$ system
(A=Co, B=Pt). The parameter $J=(V_{\rm AA}+V_{\rm BB}-2V_{\rm AB})/4$
fixes the bulk order-disorder temperature $T_0 = 1.83J/k_{\rm B}$
\cite{Binder80}. With $T_0 = 958~{\rm K}$ \cite{Berg+72} we take
$k_{\rm B}T_0/1.83=45~{\rm meV}$ as our energy unit. The parameter
$h=V_{\rm BB}-V_{\rm AA}$ controls the surface segregation of B atoms
at equilibrium. Segregation with nearly 100\% Pt at the surface
\cite{Gauthier+92} is compatible with $h\simeq4$. The parameter
$V_0=(V_{\rm AB}+V_{\rm BB})/2$ represents the average bond energy in
the L1$_2$-ordered crystal, and together with $J$ and $h$ uniquely
determines the bond energies $V_{\rm AA}$, $V_{\rm AB}$, and $V_{\rm
  BB}$. Its estimation from the cohesive energy of L1$_2$ ordered
CoPt$_3$, however, would not be appropriate, because in real materials
the total binding energy is not given by a sum of independent bond
interactions \cite{Yang+97}. Rather we note that during cluster growth
the mobility is effectively restricted to the surface, so that the
number of relevant bonds per atom is between 3 (for a diffusing
adatom) and 7 (for an atom at an edge). Parameters for a variety of
corresponding processes were calculated for Pt with DFT
\cite{Feibelman99} and yield $V_0\simeq -5$, which we adopt here.

\begin{figure}[t]
\epsfig{file=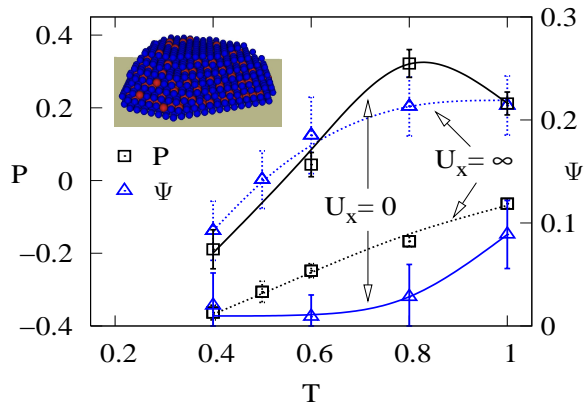,width=8cm}\vspace*{-0.2cm}
\caption{(color online) Structural anisotropy $P$ and L1$_2$ order 
  parameter $\Psi$ for clusters of size 1000 for different substrate
  temperatures, with $h=4$. The solid (dashed) lines are a guide to
  the eye for exchange barriers $U_{\rm x}=0$ ($U_{\rm x}=\infty$).
  The inset shows a typical configuration ($U_{\rm x} = 0$, $T=1)$
  before subsequent equilibration. Pt atoms are marked in blue (dark)
  and Co atoms in red.}
\label{fig:P-T}
\end{figure}\noindent

Binding to the substrate surface is represented by a weak potential
$V_{\rm s}=-5$ restricted to adatoms in the first layer. Compared to a
Pt(111) surface with three bonds of typical strength $V_0$, this
amounts to about $1/3$ of the energy of a single Pt-Pt bond.

To model the growth of the clusters, we perform kinetic Monte-Carlo
(KMC) simulations on an fcc lattice with periodic boundary conditions
in the lateral directions. Compared to previous growth simulations of
binary alloys with solid-on-solid models or immediate downward
funneling, our program is fully three-dimensional, so that for
instance diffusion on equivalent side facets is automatically
equivalent to diffusion on the top terrace. A rejectionless
continuous-time algorithm is implemented that generates cluster
configurations with statistical weights according to the underlying
master equation. The possible elementary processes are jumps of A or B
atoms to vacant nearest neighbor sites and the deposition of an atom.
The jump of an atom occurs with a rate $\nu\exp(-U/k_{\rm
  B}T)\min(1,\exp[-\Delta E/k_{\rm B}T])$, where $\Delta E$ is the
energy difference before and after the jump, $U$ is an energy barrier
and $\nu$ an attempt frequency. Diffusion measurements of Pt on
Pt(111) give $\nu\simeq5\times10^{12}\;{\rm s}^{-1}$ and $U\simeq5$.
These values are taken to be the same for all jump processes.
Deposition of atoms occurs with a rate $F=21$ monolayers per second.
The overall procedure automatically entails adatom diffusion,
deposition and re-evaporation, nucleation, step edge diffusion, and
step edge barriers. 

A typical cluster with $N=1000$ atoms is shown in the inset of
Fig.~\ref{fig:P-T}. As in the experiments \cite{Albrecht+01}, side
facets are of \{111\} and \{100\} type and the aspect ratio between
lateral and height dimensions is about 3:1.

To quantify the long-range chemical ordering we determine the average
intensity $I$ of L1$_2$ superstructure peaks and normalize it with
respect to surrogate structures, where the cluster shape is
maintained, but the atomic configuration is changed. By occupying
precisely one of the four sc-sublattices with Co atoms we obtain the
intensity $I_{\rm id}$ for perfect L1$_2$ ordering, while a random
occupation with the stoichiometric concentration yields $I_{\rm ran}$.
The order parameter is then defined as $\Psi=(I-I_{\rm ran})/(I_{\rm
  id}-I_{\rm ran})$. As can be seen from Fig.~\ref{fig:P-T}
(triangles, dashed line), $\Psi$ decreases by lowering the temperature
due to kinetic suppression of the equilibrium state, in accordance
with experiments. The extrapolated onset temperature for
L1$_2$-ordering, however, is much lower than in experiments. On the
other hand, since the deposition rate in our simulations is larger
than the experimental one, one expects a shift of calculated relative
to experimental data to higher $T$.

This problem can be resolved by allowing for direct exchange processes
between unlike neighboring atoms near the surface. As shown below,
such processes also drive the Pt surface segregation which is crucial
for producing PMA. In the model described before, only indirect
exchange processes between two atoms can take place via intermediate
states, as a result of two or more direct exchanges with vacancies.
These are, however, extremely rare since realization of the
intermediate states requires bond breaking steps with high activation
energies. Direct exchange processes are found for Co deposited on
Pt(111) over a wide temperature range 250--520~K
\cite{Gambardella+00,Lundgren+99,Santis+02}. They are especially
frequent for low-coordinated atoms on top of terraces (with
coordination 3) or at step edges (with coordinations 4 or 5).

To include these processes in the model, we allow a pair of unlike
neighboring atoms to exchange sites if one of them has a
coordination in the range 3 to 5  and the other one in the range 8
to 10. In addition to the  difference between initial and final
energies,  an increased exchange barrier $U_{\rm x} +U$ has to be
surmounted. Setting $U_{\rm x}=0$ as the opposite extreme to
$U_{\rm x}=\infty$ (no exchange processes), we see in Fig.~1 that
the onset of L1$_2$-ordering  occurs near  $T\simeq1$
(corresponding to $T=523$~K), which is in reasonable accordance
with experiments.

We now investigate the possible occurrence of a structural anisotropy
that can induce PMA. Regarding magnetic properties in Co-Pt alloys, we
adopt the viewpoint, suggested by the experiments discussed above
\cite{Nakajima+98,Shapiro+99}, that the crystalline magnetic
anisotropy originates mainly from Co-Pt bonds, which tend to align the
Co moments along the bond. PMA will therefore be traced back to a
prevalence of out-of-plane relative to in-plane Co-Pt bonds. Denoting
the corresponding total numbers of bonds in a cluster of $N$ atoms by
$n^{\rm CoPt}_\perp$ and $n^{\rm CoPt}_{\|}$, the pertinent anisotropy
is introduced as $P = (n^{\rm CoPt}_\perp - n^{\rm CoPt}_{\|})/N$. PMA
requires that $P > 0$.

In the absence of exchange processes ($U_{\rm x} = \infty$), simulated
values for $P$ turn out, however, to be negative as shown in Fig. 1.
The reason is that the clusters are relatively flat and that surface
segregation of Pt is weak. As a consequence, in a clusters' surface
there exist more Co-Pt bonds in-plane than out-of-plane, while atoms
in the inner part of the cluster give no contribution to $P$. The flat
cluster shape, on the other hand, would favor PMA if strong Pt
segregation, as realized at equilibrium, could build up. Indeed, when
including exchange processes, the Pt segregation gets strongly
enhanced and $P$ can become positive. Actually, the cluster displayed
in the inset of Fig.~1, showing pronounced Pt surface segregation, was
computed with $U_{\rm x} = 0$. As seen from the figure, for $U_{\rm x}
= 0$, $P$ changes sign near $ T = 0.58$ and reaches a maximum at an
optimum temperature $T_{\rm max} \simeq 0.8$. As in experiments
\cite{Shapiro+99}, the subsequent decrease of $P$ is accompanied by
the onset and subsequent rise of L1$_2$-ordering. 

\begin{figure}[t]
\begin{center}\hspace*{-0.5cm}
\epsfig{file=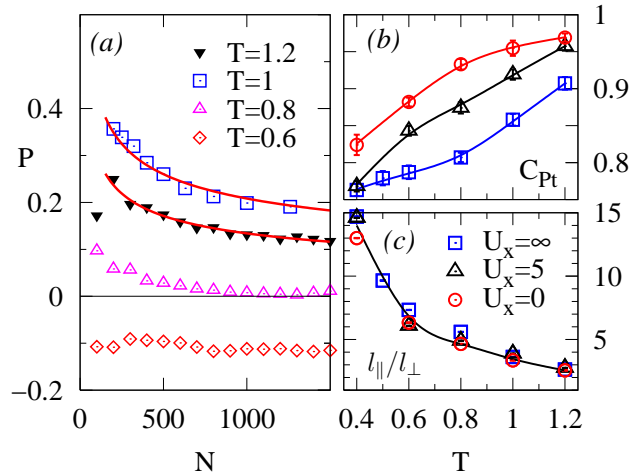,width=8.5cm}\vspace*{-0.5cm}\end{center}
\caption{\emph{(a)} Anisotropy parameter $P$ averaged over 20 clusters versus size $N$ at
  different temperatures, with $U_{\rm x} = 5$. For the two highest
  temperatures, the proportionality $ P \propto N^{-1/3}$ is well
  established, see the lines.  \emph{(b)} Pt concentration ${\rm
    C}_{\rm Pt}$ in the outer shell of the cluster for various $U_{\rm
    x}$ (symbol assignment as in \emph{(c)}).  \emph{(c)} Ratio
  $l_\parallel/l_\perp$ of cluster gyration radii as a function of
  temperature. The line is drawn as a guide to the eye.}
\label{fig:P-N}
\end{figure}

To investigate the mechanisms of PMA in more detail, we show in
Fig.~\ref{fig:P-N}a $P$ as a function of cluster size $N$ during
growth for 4 different temperatures, and for a more realistic nonzero
exchange barrier $U_{\rm x}>0$, which we set equal to the adatom
diffusion barrier, $U_{\rm x}=U=5$. Except for the two lowest
temperatures, where Pt segregation is not sufficiently strong, $P$
indeed becomes positive except for very small $N$ (not shown). Again
there exists an optimal temperature $T_{\rm max}\simeq1$, where PMA is
expected to be strongest. Changes of $P$ during subsequent
equilibration under zero flux are less then $15~\%$ of the values near
$T_{\rm max}$ and will be ignored in the following.
\noindent

The existence of an optimal temperature, however, is not directly
caused by L1$_{2}$ ordering as one might conjecture from Fig.~1.
Rather, it results from two competing effects: As shown in
Fig.~\ref{fig:P-N}b, increasing $T$ facilitates Pt segregation,
leading to larger $P$. On the other hand, increasing $T$ also drives
the cluster morphology closer to an equilibrium with less flat shapes
with smaller $P$. This effect is shown in Fig.~\ref{fig:P-N}c, where
we plot the $T$-dependence of the aspect ratio, defined by the ratio
$l_\parallel/l_\perp$ of the gyration radii of cluster sizes in the
direction parallel and perpendicular to the substrate. While the
aspect ratio does not depend on $U_{\rm x}$, the Pt segregation is
significantly enhanced for lower $U_{\rm x}$, see Fig.~\ref{fig:P-N}b.
These results can explain why there exists a temperature window for
PMA in experiments and why in films with tall rather than flat
clusters, parallel instead of perpendicular magnetic anisotropy is
favored \cite{Albrecht+02}.

Since the positive values of $P$ arise due to the surface segregation
of Pt, we expect $n_\perp^{\rm CoPt}-n_\|^{\rm CoPt}\sim N^{2/3}$, and
accordingly $P \sim N^{-1/3}$. As shown by the solid lines in
Fig.~\ref{fig:P-N}a, this behavior is well obeyed by the data sets for
$T=1$ and $T=1.2$.

We are now in the position to estimate the magnitude of the magnetic
anisotropy $E_{\rm s} = E \{ \vec \mu \; \text{in plane}\}$ - $ E
\{\vec \mu \; \text{out of plane}\}$, as expected from the data in
Fig.~\ref{fig:P-N}. $E\{\vec \mu\}$ denotes the anisotropic part of
magnetic interactions as a function of Co moments $\vec \mu$ for
saturated magnetization. Within a simple bond picture we suppose that
each Co-atom experiences a local crystalline anisotropy, which is a
sum over the corresponding twelve Co-$\alpha$ bonds in the fcc-lattice
with $\alpha = $ Co, Pt or V(vacancy), specifying the nearest neighbor
occupation. Each bond with bond-vector $\vec \delta$ contributes a
term $-A^\alpha(\vec \mu \cdot \vec \delta/|\vec \mu||\vec \delta|)^2$
to $E\{\vec \mu\}$, with anisotropy constants $A^\alpha$ to be deduced
from experiment. Note that for saturated magnetization a
nearest-neighbor contribution to the anisotropic part of dipole-dipole
interactions has exactly the same form, so that these are included
already in the coefficients $A^\alpha$. As discussed before, the most
important contribution arises from Co-Pt bonds. Anisotropy constants
for Co-Pt in multi-layer systems and wedges were measured for instance
with magnetic torque or the magneto-optical Kerr effect
\cite{Johnson+96}. From these experiments we deduce $A^{\rm CoPt}$ =
160~$\mu$eV as a representative value. Similarly, from measurements of
Co-vacuum interfaces and theoretical investigations of a freely
standing Co-monolayer \cite{Johnson+96} we estimate $A^{\rm CoV}
\simeq -67~\mu$eV. The remaining parameter is $A^{\rm CoCo}$ for which
we retain only the dipolar contribution. Using $\mu^{\rm Co} = 1.7
\mu_{\rm B}$ \cite{Menzinger+66,Shapiro+99} and $|\vec \delta| =
2.72$~\AA\ in CoPt$_3$, this yields $A^{\rm CoCo}=23~\mu{\rm eV}$.

Within this bond model it is straightforward to calculate the
anisotropy energy $E_{\rm s}$, which can be written as\vspace*{-0.8cm}

\begin{equation}
\label{energy}
\begin{split}
  E_{\rm s} & =\frac{1}{2} \sum_{\rm \alpha = Pt, Co}
(A^{\rm Co\,\alpha} - A^{\rm CoV})(n_\perp ^{\rm Co\,\alpha} -
n_{\parallel}^{\rm CoV})\\
& = K_{\rm s} N^{2/3}
\end{split}
 \end{equation}
 The last equality in (\ref{energy}) holds on account of $P \propto
 N^{-1/3}$, see Figs.~2a and 3b. Similarly, we found that ($n^{\rm
   CoCo}_\perp - n^{\rm CoCo}_{\parallel}) \propto N^{2/3}$, but this
 term related to Co-Co bonds turns out to be significantly smaller
 than the first term in (\ref{energy}) in view of our estimates for
 $A^{\rm Co\,\alpha}$. Altogether we find $K_{\rm s} \simeq 200~\mu{\rm eV}$
 at $T = 1$ and $U_{\rm x}=5$.

\begin{figure}[t]
\begin{center}\hspace*{-0.5cm}
\epsfig{file=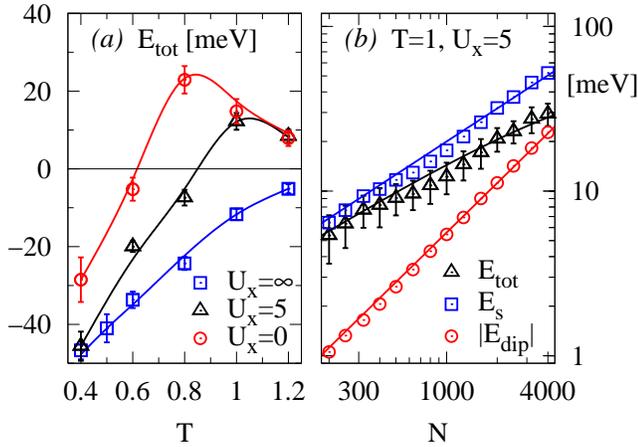,width=8.5cm}\vspace*{-0.5cm}
\end{center} \caption{
  \emph{(a)} Total magnetic anisotropy energy $E_{\rm tot}$ for
  $N=10^{3}$ in dependence of temperature for different exchange
  barriers $U_{\rm x}$ for exchange processes.  \emph{(b)} Surface and
  bulk contributions $E_{\rm s}$, $E_{\rm dip}$ in dependence of
  cluster size $N$. The corresponding straight lines in the double
  logarithmic plot have slopes 2/3 and 1.  Data for $E_{\rm tot}$ are
  fitted according to Eq.~\eqref{Etot}.}
\label{fig:ea}
\end{figure}\noindent

So far we neglected contributions to the anisotropy energy from
dipolar interactions between non-neighboring moments. These
interactions favor an orientation of the magnetization parallel to
the (111)-plane and hence give a negative bulk contribution
$E_{\rm{dip}} = - K_{\rm{dip}} N$ to the total anisotropy energy
\begin{equation}
E_{\rm tot} = K_{\rm s} N^{2/3} - K_{\rm{dip}} N \label{Etot}
\end{equation}

From calculated cluster structures, Co magnetic moments 1.7$\mu_{\rm
  B}$ and induced Pt magnetic moments 0.3$\mu_{\rm B}$
\cite{Menzinger+66,Shapiro+99} we obtain $K_{\rm {dip}} \simeq
5.6~\mu$eV at $T = 1$. This yields an $E_{\rm tot}$ that is displayed
in Fig.~3a as a function of $T$ for different $U_{\rm x}$ and $N =
1000$. An optimal temperature, where $E_{\rm tot}$ is maximum, can
clearly be identified. For example, for $U_{\rm x} = 5$, $T_{\rm max}
\simeq 1$.

In Fig.~\ref{fig:ea}b we show $E_{\rm tot}$ as a function of $N$ for
$U_{\rm x}=5$ and $T=1\simeq T_{\rm max}$, together with the surface
and dipolar contributions $E_{\rm s}\propto N^{2/3}$ and
$E_{\rm{dip}}\propto N$. The solid line for $E_{\rm tot}$ represents
Eq.~\eqref{Etot}, which predicts an optimal cluster size $N_{\rm
  opt}\simeq 13500$ with maximum $E_{\rm tot}$. The PMA is expected to
disappear for $N>N_{\rm c}\simeq 5\cdot 10^4$.

In summary, by simulating the fully three-dimensional kinetics of
binary alloy cluster growth, we have shown how PMA can result from the
combined effect of surface segregation and flat cluster morphologies.
The interplay between segregation and aspect ratio of cluster shapes
explains the occurrence of a temperature window for PMA. The effect
discussed here primarily relies on surface contributions and favors
PMA, in contrast to the dipolar form anisotropy. Stronger segregation
(e.g.\ due to lower exchange barriers) as well as smaller
adsorbate-substrate and adsorbate-vacuum surface tensions (leading to
flatter shapes) should broaden the temperature window and may guide
the search for optimized materials. We expect that for $T\simeq T_{\rm
  max}$ there exists an optimal cluster size $N_{\rm opt}$ for PMA and
a critical cluster size $N_{\rm c}=(3/2)^3N_{\rm opt}$, above which
PMA disappears. It should be interesting to test these predictions in
experiments.

We thank M.~Albrecht, M.~Maret, J.~Sch{\"a}fer and G.~Schatz for
helpful discussions and the Deutsche Forschungsgemeinschaft (SFB~513)
for financial support.

\end{document}